\documentclass[showpacs,prb,floatfix,twocolumn,amsmath]{revtex4}
\usepackage{graphicx}
\usepackage{amsmath}
\usepackage{amssymb}
\usepackage{bm}
\usepackage{dcolumn}

\begin{document}
\title{Nonlinear screening of charges induced in graphene by metal contacts}
\author{P. A. Khomyakov}
\thanks{Present address: IBM Research - Zurich;\\ E-mail: zrlpet@ch.ibm.com.}
\author{A. A. Starikov}
\author{G. Brocks}
\author{P. J. Kelly}
\affiliation{Faculty of Science and Technology and MESA$^+$ Institute for Nanotechnology, University of Twente, P.O. Box 217, 7500 AE Enschede, The Netherlands.}
\begin{abstract}
To understand the band bending caused by metal contacts, we study the potential and charge density induced in graphene in response to contact with a metal strip. We find that the screening is weak by comparison with a normal metal as a consequence of the ultra-relativistic nature of the electron spectrum near the Fermi energy. The induced potential decays with the distance from the metal contact as $x^{-1/2}$ and $x^{-1}$ for undoped and doped graphene, respectively, breaking its spatial homogeneity. In the contact region the metal contact can give rise to the formation of a $p$-$p^{\prime}$, $n$-$n^{\prime}$, $p$-$n$ junction (or with additional gating or impurity doping, even a $p$-$n$-$p^{\prime}$ junction) that contributes to the overall resistance of the graphene sample, destroying its electron-hole symmetry. Using the work functions of metal-covered graphene recently calculated by Khomyakov {\it et al.} [Phys. Rev. B {\bf 79}, 195425 (2009)] we predict the boundary potential and junction type for different metal contacts.
\end{abstract}
\date{\today}
\maketitle
\section{Introduction}
Graphene's unique electronic properties arise from the ultra-relativistic character of the electron spectrum near the Fermi energy that leads to many unusual physical effects.\cite{Novoselov:nat05,Zhang:nat05,Katsnelson:natp06,Karpan:prl07} The high mobility and low electron density (compared to normal metals) of this two-dimensional form of carbon suggest various possibilities for using single and multiple graphene sheets to make electronic devices.\cite{Novoselov:nat05,Zhang:nat05,Katsnelson:natp06,Karpan:prl07,Giovannetti:prb07,Huard:prl07,Danneau:prl08,Lee:natn08,Gorbachev:nanol08,Rotenberg:natm08} For example, one can use external gates to locally change the doping of graphene and so design $p$-$n$ or $p$-$n$-$p^{\prime}$ junctions.\cite{Huard:prl07,Cheianov:prb06} How charge inhomogeneities are then induced in graphene by the gate electrodes or by charged impurities has been studied theoretically in Refs.~\onlinecite{DiVincenzo:prb84,Katsnelson:prb06,Zhang:prl08,Fogler:prb08}.

\begin{figure}[b]
\begin{center}
\includegraphics[width=1.0\columnwidth]{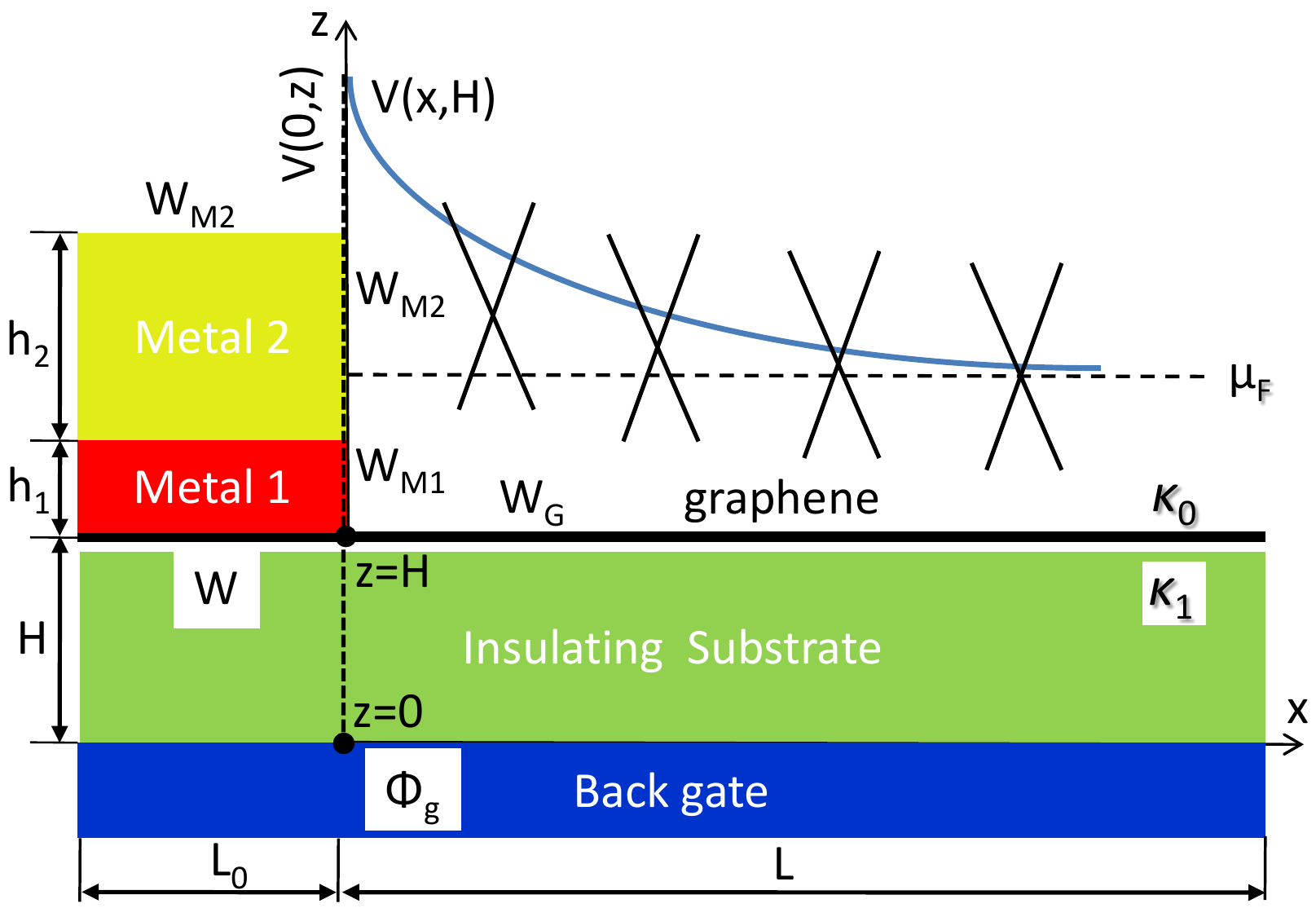}
\end{center}
\caption{(Color online) Schematic picture of a metal strip (electrode) on graphene. $V(x,z)$ is the metal-induced electrostatic potential, where $V(0,z)$ is fixed by the work functions $W_{\rm M1}$, $W_{\rm M2}$, $W_{\rm G}$ and $W$, of the bottom and top metal layers, graphene and graphene-covered metal; $h_1$, $h_2$ and $H$ are the thicknesses of the bottom and top metal electrodes and the insulating substrate, and $L_0$ and $L$ are the lateral dimensions of the electrode and the graphene sample. $\Phi_{\rm g}$ is the backgate voltage; $\kappa_{0}$ and $\kappa_{1}$ are the permittivities of the dielectric media above and below the graphene sheet. A typical experimental setup has $L_0 \lesssim L\sim 1$ $\mu$m, $H\sim 300\, {\rm nm} < L$, and $h_1 < h_2 \ll H$.
}
\label{ref:fig1}
\end{figure}

Recently, we have shown how a graphene sheet adsorbed on a metal is charged by the metal\cite{Giovannetti:prl08,Khomyakov:prb09} suggesting an alternative way of making $p$-$n$ junctions by putting weakly-interacting metal strips on graphene. The (out-of-plane) charge transfer between graphene and the metal is determined by (i) the difference between the work function of graphene and the metal surface, and (ii) the metal-graphene chemical interaction that creates an interface dipole which lowers the metal work function. Depositing a metal strip (electrode) of finite width on the graphene sheet will additionally result in an in-plane charge transfer from the graphene region covered by the metal to the ``free'' graphene supported by a dielectric, driven by the difference between the work functions of the metal, $W_{\rm M}$, of metal-covered graphene, $W$, and of free-standing graphene, $W_{\rm G}$. \cite{Khomyakov:prb09} An electrostatic potential will then be induced across the graphene sheet, leading to band bending and the formation of a $p$-$p^{\prime}$, $n$-$n^{\prime}$ or $p$-$n$ junction at the contact area as illustrated in Fig.~\ref{ref:fig1}.\cite{Huard:prb08,Lee:natn08}

In this paper we study how graphene screens charges transferred from the metal-graphene contact to ``free'' graphene, creating an electrostatic potential barrier at the contact region, see Fig.~\ref{ref:fig1}.
This problem is of fundamental importance since the physics of contacts can have a significant effect on the transport properties of an electronic device, in particular, graphene-based devices with ``invasive'' electrodes, as recently demonstrated in Ref.~\onlinecite{Huard:prb08}. A widely-used picture of a metal-graphene contact often assumes a quite sharp potential step at the contact.\cite{Tworzydlo:prl06} However, we find that the screening in graphene is strongly suppressed leading to a large-scale inhomogeneity of the electrostatic potential across the graphene sample. The contact effects can also result in charge inhomogeneities of different polarities (an abrupt $p$-$n$ junction) at the contact region. These abrupt $p$-$n$ junctions are qualitatively different from the gate-induced $p$-$n$ junctions recently reported in Refs.~\onlinecite{Huard:prl07,Cheianov:prb06,Zhang:prl08}. 

The contact phenomenon studied in the present work stems from long-range electric fields originating from charge transferred between a graphene sheet and a metal electrode. Recently, Barraza {\it et al.} have obtained a contact potential from density functional theory calculations of graphene ribbons with a length of $\leq 13.6$ nm contacted with aluminium electrodes.\cite{Barraza:prl10} To calculate the contact potential profile one applies periodic boundary conditions, i.e. periodic arrays of metal-graphene junctions. Using such results to describe a single metal-graphene junction, one assumes that the periodic images do not interact, or, in other words, that the contact potential profile is confined to a narrow region around the junction. Indeed the phenomenological model proposed in Ref.~\onlinecite{Barraza:prl10} suggests an exponential-like decay of the potential induced in graphene by a metal contact. Our results show that this does not apply to a single metal-graphene contact, where the induced potential is in fact long range. 

In Sec.~\ref{ref:sec1} we describe the Thomas-Fermi approach used to study the band banding caused by metal contacts in graphene. In Sec.~\ref{ref:sec2} we derive the induced potential and charge density for undoped, doped and gated graphene, and make material-specific predictions of the boundary potential and junction type for different metal contacts. Our conclusions are presented in Sec.~\ref{ref:sec3}.

\section{Thomas-Fermi approach}\label{ref:sec1}

We model a graphene-metal contact as shown in Fig.~\ref{ref:fig1}. Metal contacts are often applied as a bilayer; a (thin) layer of metal 1 takes care of a good adhesion to graphene, and a (thick) layer of metal 2 forms the electrode. We allow for electrostatic doping by means of a back gate electrode, and model chemical doping by a uniform background charge $\bar{\sigma}$. The resulting screening potential in graphene is calculated using density functional theory (DFT) within the Thomas-Fermi (TF) approximation. The latter has proven to have a wide range of validity for describing screening in graphene.\cite{DiVincenzo:prb84,Katsnelson:prb06,Zhang:prl08,Fogler:prb08} In TF theory the charge density $\sigma(x)$ induced in graphene for $x \geq 0$ is given in terms of the local chemical potential $\mu(x)$ as $\sigma(x)=-e D_0 \, \mu(x)\vert\mu(x)\vert/2 A$, where $D_0=0.09/$(eV$^{2}$ unit cell), $A=5.18$\,\AA$^{2}$\, is the unit cell area of graphene, and $e>0$ the elementary charge.\cite{Giovannetti:prl08}
The electrostatic potential $V(x,z)$ and the local chemical potential are related as $\mu(x) = \mu_{\rm F} - V(x)$ with $V(x) \equiv V(x,H)$ and the graphene sheet is located at $z=H$, 
where $H$ is the distance between the graphene sheet and the backgate electrode, see
Fig.~\ref{ref:fig1}. The chemical potential of the entire system, $\mu_{\rm F}$, is set to zero for undoped graphene.

Using the Poisson equation and translational symmetry in the $y$ direction, (i.e. assuming an infinitely wide graphene sample), the TF equation for the electrostatic potential induced in graphene ($x\geq0$) is
\begin{equation}\label{ref:eqn1}
\nabla^2 V(x,z) =\frac{e \left[ \sigma(x) + \bar{\sigma} \right] }{\varepsilon_{0}\kappa}\,  \delta(z-H),
\end{equation}
where $\sigma(x) = -e\lambda
 [ \mu_{\rm F} - V(x) ]\vert\mu_{\rm F} - V(x)\vert$ with
 $\lambda = D_0 /2 A$.
The effects of the substrate, the lattice potential
and of the filled band electrons are taken into account via an effective dielectric constant, $\kappa$, and a compensating charge $\bar{\sigma} = e\lambda\, \mu_{\rm F}\vert\mu_{\rm F}\vert$.\cite{Katsnelson:prb06,Ando:jpsj06} The boundary conditions for $V(x,z)$ are imposed by the potential at the contact $V(0,z)=V_{\rm C}(z)$ and at the gate $V(x,0)=V_{\rm g}(x)$.
Assuming graphene is a nearly perfect metal, i.e. $V(x) \approx {\rm const}$, which will be justified below, the boundary conditions are given by\cite{Polyanin:02,Landau:84,Shikin:prb01}
\begin{eqnarray}
V_{\rm g}(x)&=& -e\Phi_{\rm g} = {\rm const}\label{ref:eqn1a},\\
V_{\rm C}(z)&=& \left[ \left( W - W_{\rm G} \right) \frac{z}{2H} - \left( 1 - \frac{z}{H} \right) e\Phi_{\rm g} \right]\,\Theta(H-z)\nonumber\\
&+& \left[ \left( W_{\rm M} - W_{\rm G} \right) \frac{\pi}{2\beta} \right]\,\Theta(z-H),\label{ref:eqn1b}
\end{eqnarray}
where $z\geq 0$ and $\pi/2\leq\beta \leq\pi$ is the angle between the ``free'' graphene sheet and face of the metal electrode; $W_{\rm M1}$, $W_{\rm M2}$, $W_{\rm G}$ and $W$ are the work functions of the bottom and top metal layers, graphene and graphene-covered metal, respectively. $\Phi_{\rm g}$ is the backgate voltage.

Using a Green function approach, Eq.~(\ref{ref:eqn1}) can be formulated as a nonlinear integral equation\cite{Polyanin:02}
\begin{eqnarray}\label{ref:eqn2}
\hspace{-3mm} V(x) &=& V_{\rm B}(x,\Phi_{\rm g}) - e\, \int_{0}^{\infty}\frac{dx^{\prime}}{2\pi}\, \frac{\sigma(x^{\prime}) + \bar{\sigma}}{\varepsilon_{0}\kappa} \nonumber\\
&\times& {\rm ln} \left\vert\frac{(x + x^{\prime}) \sqrt{(x - x^{\prime})^2+4H^2} }{(x - x^{\prime}) \sqrt{(x + x^{\prime})^2 + 4H^2}} \right\vert ,\label{ref:eqn3}\\
\hspace{-3mm} V_{\rm B}(x,\Phi_{\rm g})  &=& \hspace{-1mm} \int_{0}^{\infty} \frac{dz^{\prime}}{\pi} \frac{4 x H\, z^{\prime} V_{\rm C} (z^{\prime})}{[x^{2} + (H-z^{\prime})^{2}][x^{2} + (H+z^{\prime})^{2}]}\label{ref:eqn4}\nonumber\\
\hspace{-3mm} &+& \hspace{-1mm} \int_{0}^{\infty} \frac{dx^{\prime}}{\pi} \frac{4 x H\, x^{\prime} V_{\rm g} (x^{\prime})}{[(x-x^{\prime})^{2} + H^{2}][(x+x^{\prime})^{2} + H^{2}]}.\hspace{1mm}
\end{eqnarray}
Using Eqs.~(\ref{ref:eqn1a}) and (\ref{ref:eqn1b}) the boundary potential term can be written as  
\begin{eqnarray}\label{ref:eqn2a}
V_{\rm B}(x, \Phi_{\rm g}) &=& \frac{2}{\pi} \left( V_{\rm B1} + V_{\rm B2} + e\Phi_{\rm g} \right)\arctan \left( \frac{2H}{x} \right) -e \Phi_{\rm g} \nonumber \\
&-& \frac{e \Phi_{\rm g} + 2 V_{\rm B1}}{2 \pi H} x\ln \left( 1 + \frac{4 H^{2}}{x^{2}} \right),
\end{eqnarray}
where the boundary potential constants $V_{\rm B1}$ and $V_{\rm B2}$ are
\begin{equation}
\label{ref:eqn5}
V_{\rm B1} =
        \frac{1}{4} (W - W_{\rm G}),\quad
V_{\rm B2} =
        \frac{\pi}{4\beta}(W_{\rm M} - W_{\rm G}),
\end{equation}
and $W$ and $W_{\rm G}$ are the work functions of the graphene-covered metal (M1) surface as calculated in Ref.~\onlinecite{Khomyakov:prb09} using the DFT within the local density approximation, and that of free-standing graphene, respectively. The influence of the electrode depends upon the geometry and the electrode work function $W_{\rm M}=W_{\rm M1}$ for $h_{2}=0$ and $W_{\rm M} = W_{\rm M2}$ for $h_{2}\gg h_{1}$. The  parameter $\beta$ of the contact geometry is $\pi/2$ for $x\ll h = h_1+h_2$ and $\pi$ for $x\gg h$. At a distance $x\sim h$ the two solutions of Eq.~(\ref{ref:eqn2}) obtained with these two values of $\beta$ serve as lower and upper bounds for the screening potential $V(x)$.

In terms of the sine Fourier transform $f(k)=\sqrt{2/\pi}\int_{0}^{\infty} dx\, f(x)\, \sin(k x)$, Eqs.~(\ref{ref:eqn2}) and (\ref{ref:eqn2a}) reduce to
\begin{eqnarray}\label{ref:eqn6}
V(k) &=& V_{\rm B}(k,\Phi_{\rm g}) - e\,  \frac{\left( 1 - e^{-2kH} \right)}{2k}\, \frac{\sigma(k) + \bar{\sigma}(k)}{\varepsilon_{0}\kappa},\\
V_{\rm B}(k,\Phi_{\rm g})&=&\sqrt{\frac{2}{\pi k^{2}}}\left[\phantom{\frac{0}{0}}\hspace{0cm} V_{\rm B1} + V_{\rm B2} +
\left(V_{\rm B1}  - V_{\rm B2} \right) e^{-2kH}
\right. \nonumber \\
&-& \left.  \left( e\Phi_{\rm g} + 2 V_{\rm B1} \right) \frac{1 - e^{-2kH}}{2kH}  \right],\label{ref:eqn6b}
\end{eqnarray}
where ${\sigma}(k)=\sqrt{2/\pi}\int_{0}^{\infty} dx\, \sigma(x)\, \sin(k x)$ and $\bar{\sigma}(k)=(2/\pi k^{2})^{1/2}\, \bar{\sigma}$.

\section{Results}\label{ref:sec2}
We have been unable to solve Eq.~(\ref{ref:eqn2}) analytically, but we have obtained a numerical solution, as well as accurate, approximate, analytical solutions. We start by studying the solution at a large distance from the metal contact assuming that $V(x) \rightarrow V_{\infty} = {\rm const}$ for $x\rightarrow\infty$.
The classical limit of a perfect metal, i.e. $V(x)=V_{\infty}$ for all $x$, implies $V(k)= (2/\pi k^{2})^{1/2} V_{\infty}$ in Eq.~(\ref{ref:eqn6}), allowing the asymptotic charge density to be written as
\begin{equation}\label{ref:eqn7}
\frac{e \left[ \sigma(x)+\bar{\sigma} \right]}{\varepsilon_{0}\kappa} \hspace{0mm} = \hspace{0mm} \frac{4}{\pi x} \left(\hspace{0mm} V_{\rm B2}\hspace{0mm} +\hspace{0mm} V_{\rm B1}
\frac{\pi x/H}{e^{\pi x/H} - 1}\hspace{0mm} -\hspace{0mm} V_{\infty} \hspace{0mm}\right)\hspace{0mm} -\hspace{0mm} \frac{e\Phi_{\rm g}}{H},\hspace{0mm}
\end{equation}
which is derived from Eqs.~(\ref{ref:eqn6}) and (\ref{ref:eqn6b}) with the use of the inverse sine Fourier transformation.
This result is valid for any two-dimensional metallic system, including multiple layers of graphene.\cite{Shikin:prb01} However, to obtain the screening potential $V(x)$ from the charge density $\sigma(x)$ one has to use the appropriate density of states. For instance, for undoped graphene, $\mu_{\rm F} = 0$ (implying $\bar{\sigma} = 0$) and the asymptotic potential is $V_{\infty}={\rm sign}(-e\Phi_{\rm g})(\varepsilon_{0}\kappa\vert\Phi_{\rm g}\vert/e\lambda H)^{1/2}$. For a single layer of graphene the screening potential can be expressed in terms of the charge density in the general form 
\begin{equation}\label{ref:eqn7b}
V(x) = \mu_{\rm F} + {\rm sign}(\sigma)\sqrt{\frac{\vert\sigma(x) \vert}{e\lambda}},
\end{equation}
where the charge density is given by Eq.~(\ref{ref:eqn7}) for given boundary potential constants ($V_{\rm B1}$, $V_{\rm B2}$), gate voltage ($\Phi_{\rm g}$), substrate thickness ($H$) and doping level ($\mu_{\rm F}$).

In the {\it ungated} limit ($H\rightarrow\infty$), the screening potential given by Eqs.~(\ref{ref:eqn7}) and (\ref{ref:eqn7b}) becomes
\begin{equation}
\label{ref:eqn8}
V(x) = \mu_{\rm F} + {\rm sign} (\sigma)
\sqrt{\left\vert \frac{V_{\rm B}\vert V_{\rm B}\vert}{x/l_{\rm s}} - \mu_{\rm F}\vert \mu_{\rm F}\vert \right\vert},
\end{equation}
where $l_{\rm s}=\hbar\upsilon/\pi\alpha\vert V_{\rm B}\vert$ is a scaling length, $V_{\rm B}=V_{\rm B1} + V_{\rm B2}$, $\hbar\upsilon$ $=\pm\epsilon_{k}/\vert {\bf k}\vert$ $=(\pi\lambda)^{-1/2}=6.05$ eV\AA\,,\cite{Giovannetti:prl08} and $\alpha=e^{2}/4\pi\varepsilon_{0}\kappa\hbar\upsilon=2.38/\kappa$ the ``fine-structure'' constant in graphene. 

\subsection{Undoped graphene}\label{ref:secug}
For undoped graphene ($\mu_{\rm F} = 0$) the induced potential decays
asymptotically as $V(x)\sim x^{-1/2}$, i.e. the screening is strongly suppressed
as compared to a normal 2D metal, where $V(x)\sim x^{-1}$. The charge density $\sigma(x)\sim V^{2}(x) \sim x^{-1}$ behaves, however, as in a 2D metal.\cite{Shikin:prb01}
\begin{table*}[!t]
\caption{$V_{\rm B}$, Eq.~(\ref{ref:eqn5}), is the boundary potential for double 
layer electrodes with $h_{1}\ll h_{2}$, and for two contact geometries with $\beta = \pi/2$ and $\pi$.
$W_{\rm M}$, $W_{\rm G}=4.48$ eV and $W$ are
the work functions calculated for close-packed surfaces of the clean metals, for free-standing
and adsorbed graphene, respectively, see Ref.~\onlinecite{Khomyakov:prb09}. The doping sign of
adsorbed and ``free'' graphene corresponds, respectively, to the sign of
$W-W_{\rm G}$ and $V_{\rm B}=V_{\rm B1} + V_{\rm B2}$.}
\begin{ruledtabular}
\begin{tabular}{lrrrrrrrrrrrr}
                       & Ti   &  Ni   & Co   & Pd   & Al   &  Ag  & Cu   & Au   & Pt   &  Ti/Pd & Ti/Au & Ti/Al \\
\hline
 $(W_{\rm M}-W_{\rm G})\sim V_{\rm B2}$, (eV)  & 0.08 & 0.99  & 0.96 & 1.19 &-0.26 & 0.44 & 0.74 & 1.06 & 1.65 &  1.19 & 1.06 & -0.26 \\
 $(W-W_{\rm G})\sim V_{\rm B1}$, (eV)   &-0.31 &-0.82  &-0.70 &-0.45 &-0.44 &-0.24 &-0.08 & 0.26 & 0.39 &-0.31   &-0.31 &-0.31  \\
 $V_{\rm B}$ (eV), $\beta = \pi/2$  &-0.04 & 0.29  & 0.31 & 0.48 &-0.24 & 0.16 & 0.35 & 0.60 & 0.92 &  0.52  & 0.45 &-0.21  \\
 $V_{\rm B}$ (eV), $\beta = \pi$   &-0.06 & 0.04  & 0.07 & 0.19 &-0.18 & 0.05 & 0.17 & 0.33 & 0.51 &  0.22  & 0.19 &-0.14  \\
  {\rm junction type}  & $n$$-$$n^{\prime}$  & $n$$-$$p$   & $n$$-$$p$  & $n$$-$$p$ & $n$$-$$n^{\prime}$  & $n$$-$$p$  & $n$$-$$p$  & $p$$-$$p^{\prime}$  & $p$$-$$p^{\prime}$  &  $n$$-$$p$   &  $n$$-$$p$ &  $n$$-$$n^{\prime}$
\label{ref:tab1}
\end{tabular}
\end{ruledtabular}
\vspace{0.mm}
\end{table*}
\begin{figure}[!tpb]
\begin{center}
\includegraphics[width=1.0\columnwidth]{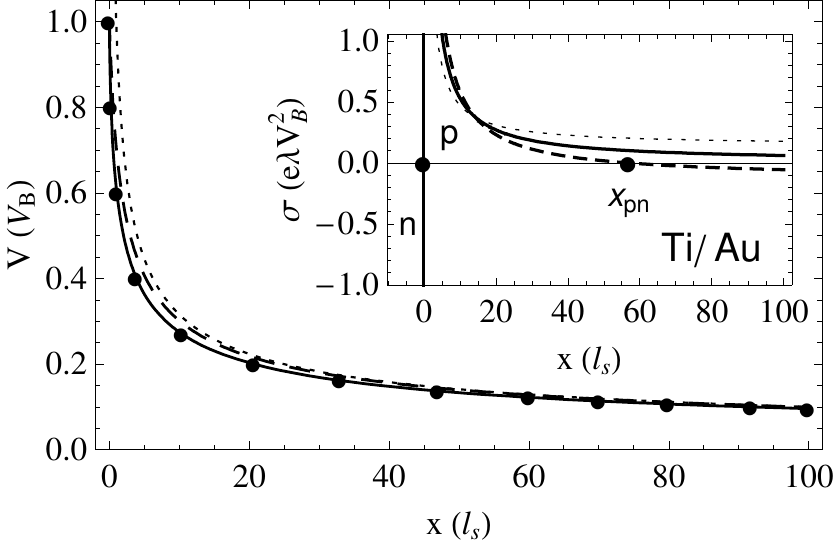}
\end{center}
\vspace{0mm}
\caption{The screening potential for undoped graphene: numerically exact solution (solid line), variational solution Eq.~(\ref{ref:eqn9}) (bold dots), $V(x)=V_{\rm B}\, (x/l_{\rm s})^{-1/2}$ (dotted line) and  $V(x)=V_{\rm B}\, (1 + x/l_{\rm s})^{-1/2}$ (dashed line).
Inset: the charge density $\sigma(x) = e\lambda V(x)\vert V(x)\vert$, Eq.~(\ref{ref:eqn7}), for the Ti/Au electrode for  $\Phi_{\rm g}=-15$ V (dashed line), $0$ V (solid line) and $15$ V (dotted line); $V_{\rm B}=0.19$ eV, $\kappa=2.5$, $\lambda = 8.69\times 10^{-3}$ (eV \AA)$^{-2}$, $H=300$ nm, $l_{\rm s}\sim 1$\,nm.
}\label{ref:fig2}
\vspace{0mm}
\end{figure}

Numerical solution of Eq.~(\ref{ref:eqn2}) shows that Eq.~(\ref{ref:eqn8}) is quite accurate for $x\gg l_{\rm s}$. A simple interpolation expression for the screening potential of undoped graphene ($\mu_{\rm F} = 0$) can be obtained by replacing $x/l_{\rm s}$ with $1+x/l_{\rm s}$ in Eq.~(\ref{ref:eqn8}). The screening potential then satisfies the correct boundary conditions both at $x=0$ and at $x=\infty$, and is close to the exact solution also for intermediate $x$, see Fig.~\ref{ref:fig2}.\cite{footnote} The best variational solution we found is 
\begin{equation}\label{ref:eqn9}
\hspace{0mm} V(x) \hspace{0mm} \approx \hspace{0mm} \frac{V_{\rm B}}{\left(\sqrt{x/l_{\rm s} + \beta_{2}^{2}} + \beta_{1} - \beta_{2}\right)^{1/2} \hspace{0mm} \left( x/l_{\rm s} + \beta_{1}^{-2} \right)^{1/4} },\hspace{0mm}
\end{equation}
with $\beta_1=0.915$ and $\beta_2=0.128$. The difference between the numerical and variational solutions is $\lesssim 1.4\%$ for $x\lesssim 0.1\,l_{\rm s}$ and $\lesssim 0.2\%$ for $x\gtrsim 0.1\,l_{\rm s}$.
\begin{figure}[b]
\begin{center}
\includegraphics[width=1.0\columnwidth]{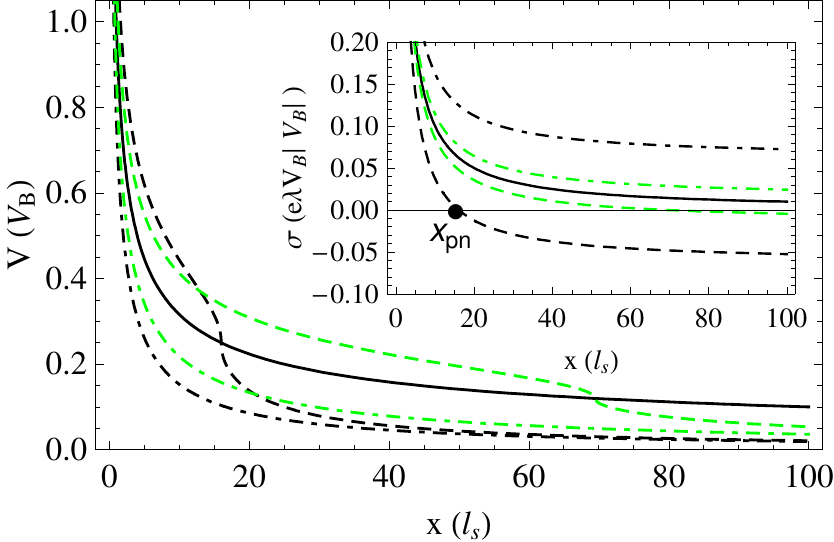}
\end{center}
\vspace{0mm}
\caption{(Color online) The screening potential for doped graphene, Eq.~(\ref{ref:eqn8}): $\mu_{\rm F}/V_{\rm B}=-0.12$ (dot-dashed green/gray line) and $-0.25$ (dot-dashed black line); $\mu_{\rm F}/V_{\rm B}=0.12$ (dashed green/gray line) and $0.25$ (dashed black line). As a reference we also show the screening potential for undoped graphene, $\mu_{\rm F} = 0$, (solid line). Inset: the charge density for doped graphene. 
}\label{ref:fig3}
\vspace{0mm}
\end{figure}

From the general criterion for validity of the semiclassical approximation,  $\vert d\lambda(x)/dx \vert/2\pi\ll 1$, where $\lambda(x)$ is the de Broglie wavelength,\cite{Landau:77} one can derive a condition defining the range of validity of the potential obtained from the TF model: $\vert  dV(x)/dx\vert\ll V^{2}(x)/\hbar\upsilon$. The latter condition also implies that the in-plane electric field is small compared to the field perpendicular to the graphene sheet, $E_x \ll E_z$ (as $E_{z}\sim\sigma\sim V^{2}$), which completes the proof that graphene behaves as a nearly perfect metal within the TF approximation. Using Eq.~(\ref{ref:eqn9}), and taking a typical boundary potential $V_{\rm B}\sim 0.5$ eV (Table~\ref{ref:tab1}) and effective dielectric constant $\kappa\sim (\kappa_{0} + \kappa_{1})/2\sim 2.5$ for graphene on a SiO$_2$ substrate,\cite{Ando:jpsj06} we find that the TF theory is valid for $x\gg a$ with $a$ the lattice parameter of graphene. Using high $\kappa$ substrates will weaken this condition, making the TF results valid up to the immediate proximity of the metal contact.

Recently we described how depositing graphene on a metal surface leads to charge transfer between the metal and graphene. \cite{Giovannetti:prl08,Khomyakov:prb09} The graphene device shown in Fig.~\ref{ref:fig1} consists of two regions, metal-covered graphene ($x<0$) and ``free'' graphene ($x>0$). The sign and level of doping of metal-covered graphene is fixed by the first metal layer (M1), $\Delta E_{\rm F} = W - W_{\rm G}$.\cite{Giovannetti:prl08,Khomyakov:prb09} The doping of graphene near the contact, $a\ll x\ll H/\pi$, is determined by the boundary potential $V_{\rm B}$, see Eq.~(\ref{ref:eqn9}), which depends on the work function of the graphene-covered metal (M1) and the top metal layer (M2), see Eq.~(\ref{ref:eqn5}). If ${\rm sign}(V_{\rm B})\neq {\rm sign}(\Delta E_{\rm F})$ an abrupt $p$-$n$ junction should form close to the contact at $x\sim a$, see inset in Fig.~\ref{ref:fig2}. \cite{footnote3} $V_{\rm B}$ and $\Delta E_{\rm F}$ having the same sign but a different size results in $p$-$p^{\prime}$ or $n$-$n^{\prime}$ junctions. Such junctions break the electron-hole symmetry in graphene, giving rise to an asymmetric resistance as a function of the gate voltage.\cite{Huard:prb08,Cayssol:prb09,Farmer:nl09}

At a distance $x\gg H/\pi$ the induced charge is determined by the top metal layer (``Metal 2''), $V_{\rm B2}\sim W_{\rm M2} - W_{\rm G}$, whereas far from the contact, $x \gg \max(L_{0},h)$, graphene is unaffected by the metal electrode. Using the work functions calculated in Ref.~\onlinecite{Khomyakov:prb09}, the boundary potential and junction type expected for different metal contacts are listed in Table~\ref{ref:tab1}, assuming the metal surfaces are clean. Because of the sensitivity of work functions to surface contamination, the contact potentials may be different for experiments that are not performed under ultrahigh vacuum conditions.\cite{footnote2}

\subsection{Doped graphene}
Real graphene samples are often doped with some form of charged impurities. Eq.~(\ref{ref:eqn8}) shows that the screening potential is relatively unchanged by doping for a graphene region where $\vert V(x)\vert \gg \vert\mu_{\rm F}\vert$, and asymptotically behaves as $V_{\rm B}\vert V_{\rm B}\vert l_{\rm s}/2\vert \mu_{\rm F}\vert x$ for $\vert V(x)\vert \ll \vert\mu_{\rm F}\vert$, see Fig.~\ref{ref:fig3}, i.e. the screening is enhanced by doping. 

Eq.~(\ref{ref:eqn8}) reveals an interesting effect related to the sign of doping.
The screening potential is different for graphene doped with electrons and holes due to a $p$-$n$ junction formation at $x_{pn}=l_{s}\,(\mu_{\rm F}/V_{\rm B})^{-2}$ for $\mu_{\rm F}/V_{\rm B}>0$ as shown in Fig.~\ref{ref:fig3}. 
 The formation of the {\it p}-{\it n} junction for $\mu_{\rm F}/V_{\rm B}>0$ is caused by the competition between the charges of opposite sign induced by the metal contact and the charged impurities, respectively. The screening charge and potential for $x<x_{pn}$ ($x > x_{pn}$) are constrained by the metal contact (the charged impurities). The screening at $x\sim x_{pn}$ is then suppressed, which gives rise to increase of the screening potential at the {\it p}-{\it n} junction as compared to the case of $\mu_{\rm F}/V_{\rm B}<0$.\cite{footnote0}  For $\mu_{\rm F}/V_{\rm B}<0$ the charges induced in graphene by the metal contact and the charged impurities are of the same sign so the overall effect leads to enhancement of the screening, reducing the induced potential.  
The asymmetry reaches its maximum at a critical doping level $\vert\mu_{\rm F}\vert\sim 0.1\, \vert V_{\rm B}\vert$ and vanishes upon increasing (decreasing) $\mu_{\rm F}$, i.e. for $\vert\mu_{\rm F}\vert\gg 0.1\, \vert V_{\rm B}\vert$ ($\vert\mu_{\rm F}\vert\ll 0.1\, \vert V_{\rm B}\vert$). This effect should be observable in transport measurements where the impurity doping is gradually changed from $n$- to $p$-type.\cite{Farmer:nl09}

\subsection{Gated graphene}
One can also create $p$-$n$ junctions by applying a gate voltage (Fig.~\ref{ref:fig1}). \cite{Cheianov:prb06,Zhang:prl08,Fogler:prb08} Eq.~(\ref{ref:eqn7}) shows that the first effect of gating graphene is to cause a constant shift of the induced charge density.
Secondly, the charge depletion width can be varied by changing the gate voltage since $V_{\infty}\sim \vert\Phi_{\rm g}\vert^{1/2}$. This accounts for the electric field effect on the work function of graphene.\cite{Yu:nl09}
Thirdly, a gate voltage applied to graphene creates qualitatively different junctions ($n$-$n^{\prime}$, $p$-$p^{\prime}$, $p$-$n$ or $n$-$p$) at the contact region depending on the sign of the voltage. This will then lead to an asymmetry in the transport characteristics of the graphene device rather similar to the contact-induced abrupt junctions discussed in the last two paragraphs of Sec.~\ref{ref:secug}.

The point $x_{pn}$ that separates $p$ and $n$-doped regions of graphene, is given by  $\sigma(x_{pn})=0$. Eq.~(\ref{ref:eqn7}) yields $\pi x_{pn}/4H\approx (V_{\rm B} - V_{\infty})/e\Phi_{\rm g}$. The screening potential near the transition point can be calculated using TF theory. The semiclassical description, however, breaks down exactly at $x_{pn}$ since there are no screening charges at this point. The quantum corrections to the TF theory are, nevertheless, relatively small in graphene. \cite{DiVincenzo:prb84,Zhang:prl08,Fogler:prb08}

According to our analysis, gating graphene (or impurity doping) can create an $n$-$p$-$n^{\prime}$ junction for some metal electrodes (Ni, Co, Pd, Ag, Cu, Ti/Pd and Ti/Au), see Table~\ref{ref:tab1} and Fig.~\ref{ref:fig2}. This is possible because the sign of the graphene doping beneath the electrode, near the contact and far from the contact is independently fixed by the bottom metal layer, the contact potential and the gate voltage (impurity doping), respectively.

\section{Conclusions}\label{ref:sec3}
We have studied the electrostatic barrier formed in graphene in response to a metal strip in contact with the graphene sheet. By comparison with conventional metals, the screening in graphene is strongly suppressed: the induced electrostatic potential decays weakly with the distance from the metal contact as $V(x)\sim x^{-1/2}$ and $\sim x^{-1}$ for undoped and doped graphene, respectively. This leads to a substantial space charge region in graphene, breaking its spatial homogeneity. The latter has been recently observed by scanning photocurrent microscopy,\cite{Lee:natn08} and might also be seen for graphene in the quantum Hall regime.\cite{Shikin:prb01,Giesbers:apl08} The contact effects also result in the formation of a $p$-$p^{\prime}$, $n$-$n^{\prime}$ or $p$-$n$ junction at the near contact area that breaks the electron-hole symmetry and contributes to the contact resistance.\cite{Huard:prb08,Farmer:nl09,Cayssol:prb09,Yu:nl09,Giesbers:apl08} 
We predict that $n$-$p$-$n^{\prime}$ junctions can be realized by gating graphene or by impurity doping in combination with specific metal electrodes. 

This work was financially supported by the ``Nederlandse Organisatie voor Wetenschappelijk Onderzoek (NWO)'' via
the research programs of ``Chemische Wetenschappen (CW)'' and the ``Stichting voor Fundamenteel Onderzoek der Materie (FOM)''.


\vspace{0mm}

\begin{thebibliography}{999}

\bibitem{Novoselov:nat05} K. S. Novoselov, A. K. Geim, S. V. Morozov, D. Jiang, M. I. Katsnelson, I. V. Grigorieva, S. V. Dubonos, and  A. A. Firsov, Nature {\bf 438}, 197 (2005).

\bibitem{Zhang:nat05}  Y. B. Zhang, Y.-W. Tan, H. L. Stormer, and P. Kim, Nature {\bf 438}, 201 (2005).

\bibitem{Katsnelson:natp06} M. I. Katsnelson, K. S. Novoselov, and A. K. Geim, Nat. Phys. {\bf 2}, 620 (2006).

\bibitem{Karpan:prl07} V. M. Karpan, G. Giovannetti, P. A. Khomyakov, M. Talanana, A. A. Starikov, M. Zwierzycki, J. van den Brink, G. Brocks, and P. J. Kelly, Phys. Rev. Lett. {\bf 99}, 176602 (2007).

\bibitem{Giovannetti:prb07} G. Giovannetti, P. A. Khomyakov, G. Brocks, P. J. Kelly, and J. van den Brink, Phys. Rev. B {\bf 76}, 073103 (2007).

\bibitem{Huard:prl07} B. Huard, J. A. Sulpizio, N. Stander, K. Todd, B. Yang, and D. Goldhaber-Gordon, Phys. Rev. Lett. {\bf 98}, 236803 (2007).

\bibitem{Danneau:prl08}  R. Danneau, F. Wu, M. F. Craciun, S. Russo, M. Y. Tomi, J. Salmilehto, A. F. Morpurgo, and P. J. Hakonen, Phys. Rev. Lett. {\bf 100}, 196802 (2008).

\bibitem{Lee:natn08} E. J. H. Lee, K. Balasubramanian, R. T. Weitz, M. Burghard, and K. Kern, Nat. Nanotech. {\bf 3}, 486 (2008);
T. Mueller, F. Xia, M. Freitag, J. Tsang, and Ph. Avouris, Phys. Rev. B {\bf 79}, 245430 (2009).

\bibitem{Gorbachev:nanol08}  R. V. Gorbachev, A. S. Mayorov, A. K. Savchenko, D. W. Horsell, and F. Guinea, Nano Lett. {\bf 8}, 1995 (2008).

\bibitem{Rotenberg:natm08} E. Rotenberg, A. Bostwick, T. Ohta, J. L. McChesney, T. Seyller, and  K. Horn, Nat. Materials {\bf 7}, 258 (2008).

\bibitem{Cheianov:prb06} V. V. Cheianov and V. I. Fal'ko, Phys. Rev. B {\bf 74}, 041403 (2006).

\bibitem{DiVincenzo:prb84} D. P. DiVincenzo and E. J. Mele, Phys. Rev. B {\bf 29}, 1685 (1984).

\bibitem{Katsnelson:prb06} M. I. Katsnelson, Phys. Rev. B {\bf 74}, 201401 (2006).

\bibitem{Zhang:prl08} L. M. Zhang and M. M. Fogler, Phys. Rev. Lett. {\bf 100}, 116804 (2008).

\bibitem{Fogler:prb08} M. M. Fogler, D. S. Novikov, L. I. Glazman, and B. I. Shklovskii, Phys. Rev. B {\bf 77}, 075420 (2008).

\bibitem{Giovannetti:prl08}  G. Giovannetti, P. A. Khomyakov, G. Brocks, V. M. Karpan, J. van den Brink, and P. J. Kelly, Phys. Rev. Lett. {\bf 101}, 026803 (2008).

\bibitem{Khomyakov:prb09} P. A. Khomyakov, G. Giovannetti, P. C. Rusu, G. Brocks, J. van den Brink, and P. J. Kelly, Phys. Rev. B {\bf 79}, 195425 (2009).

\bibitem{Huard:prb08} B. Huard, N. Stander, J. A. Sulpizio, and D. Goldhaber-Gordon, Phys. Rev. B {\bf 78}, 121402(R) (2008).

\bibitem{Tworzydlo:prl06} J. Tworzydlo, B. Trauzettel, M. Titov, A. Rycerz, and C. W. J. Beenakker, Phys. Rev. Lett. {\bf 96}, 246802 (2006).

\bibitem{Barraza:prl10} S. Barraza-Lopez, M. Vanevi\'c, M. Kindermann, and M. Y. Chou, Phys. Rev. Lett. {\bf 104}, 076807 (2010).

\bibitem{Ando:jpsj06} T. Ando, J. of Phys. Soc. of Japan {\bf 75}, 074716 (2006).


\bibitem{Polyanin:02} A. D. Polyanin, ``Handbook of Linear Partial Differential Equations for Engineers and Scientists''
(Chapman \& Hall/CRC, 2002), Eqs.~7.2.2-5 and 7.2.2-12.

\bibitem{Landau:84} L. D. Landau and E. M. Lifshits, "Electrodynamics of Continuous Media" (Pergamon Press, 1984).

\bibitem{Shikin:prb01} V. B. Shikin, Phys. Rev. B {\bf 64}, 245335 (2001).

\bibitem{footnote} In principle, a similar interpolation procedure can be applied for the screening potential of doped graphene by replacing $x/l_{s}$ with $( x_{0} + x )/l_{s}$ in Eq.~(\ref{ref:eqn8}), where $x_{0} = \left[ \xi \vert\xi\vert + (1 - \xi) \vert 1 - \xi \vert \right]^{-1}$ and $\xi = \mu_{\rm F}/V_{\rm B}$.

\bibitem{Landau:77} L. D. Landau and E. M. Lifshits, "Quantum Mechanics: Non-Relativistic Theory" (Pergamon Press, 1977).

\bibitem{footnote3} The potential near the contact, $x$ and $\vert z-H\vert \sim a$, will be very sensitive to the atomic structure of the metal (M1) electrode. However, this will not affect the asymptotic behavior of the potential, $x^{-1/2}$, at $x\gg a$ since the contribution to the TF screening potential from this small region, $\delta z\sim a$, is of higher order, $\sim x^{-3/2}$, see Ref.~\onlinecite{footnote2}.

\bibitem{Farmer:nl09} D. B. Farmer, R. Golizadeh-Mojarad, V. Perebeinos, Y.-M. Lin, G. S. Tulevski, J. C. Tsang, and Ph. Avouris, Nano Lett. {\bf 9}, 388 (2009).
    
\bibitem{Cayssol:prb09} R. Golizadeh-Mojarad and S. Datta, Phys. Rev. B {\bf 79}, 085410 (2009); J. Cayssol, B. Huard, and D. Goldhaber-Gordon, Phys. Rev. B, {\bf 79}, 075428 (2009).

\bibitem{footnote2} In some experiments the charged impurities might screen the potential such that the region where the contact potential $V_{\rm C}(z)$ is non-zero gets narrowed down to a small area near the contact, $\vert z-H \vert\lesssim \delta z$. Since the boundary potential then is $V_{\rm B}(x,0)=2V_{\rm B}\arctan(\delta z/x)/\pi$, the screening potential $V(x)=V_{\rm B} [x(1+x^{2}/\delta z^{2})/l_{\rm s}]^{-1/2}$ behaves as $x^{-1/2}$ for $a\ll x\lesssim \delta z$ and $x^{-3/2}$ for $x\gg \delta z$.

\bibitem{footnote0} We note that the screening for $\mu_{\rm F}/V_{\rm B} > 0$ is even less efficient as compared to undoped graphene ($\mu_{\rm F}=0$) in the region of $a\ll x<x_{pn}$, see Fig.~\ref{ref:fig3}.

\bibitem{Yu:nl09} Y.-J. Yu, Y. Zhao, S. Ryu, L. E. Brus, K. S. Kim, and P. Kim, Nano Lett. {\bf 9}, 3430 (2009).

\bibitem{Giesbers:apl08} A. J. M. Giesbers, G. Rietveld, E. Houtzager, U. Zeitler, R. Yang, K. S. Novoselov, A. K. Geim, and J. C. Maan, Appl. Phys. Lett. {\bf 93}, 222109 (2008); P. Blake, R. Yang, S. V. Morozov, F. Schedin, L. A. Ponomarenko, A. A. Zhukov, R. R. Nair, I. V. Grigorieva, K. S. Novoselov, and A. K. Geim, Solid State Comm. {\bf 149},  1068 (2009); S. Russo, M. F. Craciun,  M. Yamamoto, A. F. Morpurgo, and S. Tarucha, Physica E {\bf 42}, 677 (2010).

\end{thebibliography}
\end{document}